\documentstyle[prb,twocolumn,aps]{revtex}
\input{epsf}
\begin{document}
\draft
\twocolumn[\hsize\textwidth\columnwidth\hsize\csname @twocolumnfalse\endcsname
\title{Disordered Hubbard Model with Attraction:\\
Coupling Energy of Cooper Pairs in Small Clusters}

\author{Jos\'e Lages$^{(a)}$, Giuliano Benenti$^{(a,b,c)}$,
and Dima L. Shepelyansky$^{(a)}$}

\address{$^{(a)}$Laboratoire de Physique Quantique, UMR 5626 du CNRS, 
Universit\'e Paul Sabatier, 31062 Toulouse Cedex 4, France}
\address{$^{(b)}$International Center for the Study of Dynamical 
Systems, \\ 
Universit\`a degli Studi dell'Insubria, via Valleggio 11, 22100 Como, Italy}  
\address{$^{(c)}$Istituto Nazionale di Fisica della Materia, 
Unit\`a di Milano, via Celoria 16, 20133 Milano, Italy} 

\date{\today}

\maketitle

\begin{abstract}
We generalize the Cooper problem to the case of many interacting 
particles in the vicinity of the Fermi level in the presence of 
disorder. On the basis of this approach we study numerically the 
variation of the pair coupling energy in small clusters as a 
function of disorder. We show that the Cooper pair energy is 
strongly enhanced by disorder, which at the same time leads 
to the localization of pairs. 
\end{abstract}
\pacs{PACS numbers:  72.15.Rn, 71.30+h, 74.20.-z}
\vskip1pc]

\narrowtext

\section{Introduction} 
\label{intro} 

The superconductor-insulator transition (SIT) in disordered 
films of metals has attracted widespread interest in recent 
years \cite{exp}. The SIT, driven by adjusting some tuning  
parameter such as film thickness or magnetic field strength,   
is particularly interesting in two dimensions (2D), where both 
superconductivity and metallic behavior are marginal.     

In the composite bosons model, Cooper pairs are treated as 
point-like charge $2e$ bosons. In this picture, 
the superconducting state displays a quantum phase transition 
to an insulating state characterized by a quenching of the 
condensate of composite bosons \cite{fisher}. In this scenario, 
the SIT is caused by the loss of phase coherence between the 
pairs in different parts of the sample, while the magnitude of 
the pairing gap remains finite. Numerical studies support
this scenario \cite{erik}.

However, a few relevant experimental aspects of the SIT 
\cite{exp} seem to be beyond the scope of the composite 
fermions theory. 
It is therefore highly desirable to study models where 
the fermionic nature of charge carriers is not rubbed out 
from the beginning. The Quantum Monte Carlo studies of the 
disordered attractive Hubbard model in two dimensions 
have supported the possibility of a disorder driven 
superconductor to insulator quantum phase transition 
\cite{trivedi}. At the same time the mean field approach 
within the Bogoliubov-de Gennes framework has shown that 
also space fluctuations of the pairing amplitude should be 
taken into account in order to give a full picture of 
the SIT \cite{trivedi2}.  

In parallel a growing interest has been devoted to the 
question of what is the coupling energy of pairs placed 
in small superconducting grains, with the average level 
spacing of the same order as the superconducting gap  
\cite{ralph,delft,ambegaokar,larkin}.  
Also the pair properties in small size samples may be 
related to their properties in the localized phase, 
where the pair motion is bounded inside the localization 
domain.  

In this paper we study numerically the properties of 
Cooper pairs in small two-dimensional clusters with 
disorder. We take the Hubbard attractive interaction 
between fermionic particles with spin $1/2$ which move 
in a two-dimensional Anderson lattice. 
Following the approach introduced by Cooper \cite{cooper}, 
we consider some part of the particles below the Fermi sea  
as frozen, while the remaining particles, in the direct 
vicinity of the Fermi level, can move and interact in the 
presence of disorder. 
Recently, such generalized Cooper problem has been considered 
\cite{jose} for the case of two particles in a disordered 
potential. Here, we further develop this approach for the 
case of many interacting Cooper pairs, that allows us to 
study the case of finite particle density.  

Our numerical studies allow us to determine the dependence 
of the Cooper pair coupling energy on the strength of disorder. 
They show that this coupling can be strongly increased by 
disorder, which however leads to localization of pairs. 
In the regime of weak disorder the pairs are delocalized 
but their coupling energy is significantly reduced 
compared to the localized phase. 

The paper is organized as follows: In Section \ref{model} 
we introduce the attractive Hubbard model and discuss the 
numerical method used to study the case of a finite particle 
density. In Section \ref{gap} we determine the Cooper pair 
coupling energy (pairing gap) and investigate its dependence 
on the strength of the disorder. 
In Section \ref{localization} we study the disorder-induced 
pair localization and compare the results with the case of 
noninteracting particles. 
In Section \ref{pair} we study the behavior of the superconducting 
order parameter, obtained from the pairing correlation function. 
The conclusions are presented in Section \ref{conc}. 

\section{Model and Numerical Method} 
\label{model} 

We study a disordered square lattice with $N$ fermions 
on $L^2$ sites. The Hamiltonian is defined by  
\begin{equation} 
\label{hamiltonian} 
H=-V\sum_{<{\bf ij}>\sigma} 
c_{{\bf i}\sigma}^\dagger c_{{\bf j}\sigma} 
+\sum_{{\bf i}\sigma} \epsilon_{\bf i} 
n_{{\bf i}\sigma} +U \sum_{\bf i} n_{{\bf i} \uparrow} 
n_{{\bf i} \downarrow}, 
\end{equation} 
where $c_{{\bf i}\sigma}^\dagger$ ($c_{{\bf i}\sigma}$) 
creates (destroys) an electron at site ${\bf i}$ with spin $\sigma$, 
$n_{{\bf i}\sigma}=c_{{\bf i}\sigma}^{\dagger}
c_{{\bf i}\sigma}$ is the corresponding 
occupation number,     
the hopping term $V$ between nearest neighbors lattice sites 
characterizes the kinetic energy, 
the site energies $\epsilon_{\bf i}$ are taken from a box 
distribution over $[-W/2,W/2]$, 
$U$ measures the strength of the Hubbard attraction ($U<0$),   
and periodic boundary conditions are taken in both directions. 
We restrict our numerical investigations to the subspace with 
$S_z=0$ for even $N$ ($N/2$ spins up and $N/2$ spins down) and 
$S_z=1/2$ for odd $N$.   

The model (\ref{hamiltonian}) at $U=0$ reduces to the one body 
Anderson model, giving localized states in two dimensions at the 
thermodynamic limit \cite{abrahams}. At $W=0$ one gets the 
clean attractive Hubbard model, which in 2D shows a Kosterlitz-Thouless 
transition to a superconducting state with power-law decay of the 
pairing correlations \cite{dagotto}.   
  
We study numerically the model (\ref{hamiltonian}) for a finite 
density of interacting quasiparticles above the frozen Fermi sea:  
\newline\noindent 
(i) Single particle eigenvalues $E_\alpha$ and eigenstates 
(orbitals) $\phi_\alpha({\bf i})$ ($\alpha=1,...,L^2$) at $U=0$ are 
obtained via numerical diagonalization of the Anderson 
Hamiltonian.      
\newline\noindent 
(ii) The Hamiltonian (\ref{hamiltonian}) is written in 
the orbital basis: 
\begin{equation} 
\label{hamobb} 
H=\sum_{\alpha\sigma} 
E_\alpha d_{\alpha\sigma}^\dagger d_{\alpha\sigma}+ 
U\sum_{\alpha\beta\gamma\delta} Q_{\alpha\beta}^{\gamma\delta}
d_{\alpha\uparrow}^\dagger d_{\beta\downarrow}^\dagger  
d_{\delta\downarrow} d_{\gamma\uparrow}, 
\end{equation} 
with $d_{\alpha\sigma}^\dagger=\sum_{\bf i}\phi_\alpha({\bf i}) 
c_{{\bf i}\sigma}^\dagger$, and 
transition matrix elements  
\begin{equation}   
Q_{\alpha\beta}^{\gamma\delta}= 
\sum_{\bf i} \phi_\alpha({\bf i})\phi_\beta({\bf i})
\phi_\gamma({\bf i})\phi_\delta({\bf i}).    
\end{equation} 
\newline\noindent 
(iii) The Fermi sea is introduced by restricting the sums in 
(\ref{hamobb}) 
to orbitals with energies above the Fermi energy $E_{m_F}$: 
$\alpha,\beta,\gamma,\delta> m_F$. We consider a filling factor 
$\nu=m_F/L^2=1/4$ (corresponding to $2 m_F$ frozen electrons due to 
spin degeneracy) and a finite density of $N$ interacting quasiparticles 
above the Fermi level.      
\newline\noindent 
(iv) The Slater determinants basis, built from the single particle 
orbitals $\phi_\alpha$, is energetically cut-off by means of the   
condition $\sum_{i=1}^{N} (m_i-m_F)\leq M$, with $m_i$ orbital index 
for the $i$-th quasiparticle ($m_i>m_F$). Such a rule gives an 
effective phonon frequency $\omega_D\propto M/L^2$.    
\newline\noindent 
(v) The ground state of this truncated Hamiltonian is found via 
the Lanczos algorithm \cite{sorensen}.  

In our numerical simulations we considered up 
to $N=8$ interacting quasiparticles in up to $M=20$ 
orbitals. We checked that results are qualitatively  
similar under variation of the cut-off orbital $M$. 
In the following Sections we present data for 
$U=-4V$, averaged over $N_R=100$ disorder realizations.  

\section{Pairing Gap} 
\label{gap} 

In order to compute the pairing energy, we first compute 
the total $N$-body pairing energy as 
\begin{equation} 
E_p(N)=E_g(U=0)-E_g(U),  
\end{equation} 
with $E_g(U)$ many body ground state for an attractive 
interaction $U$.   

\begin{figure}
\epsfxsize=8cm
\epsffile{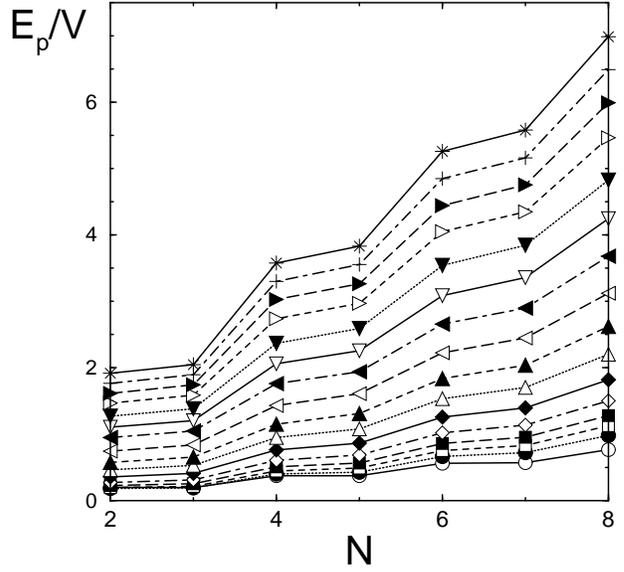}
\caption{Dependence of the $N$-body pairing energy 
$E_p$ on the number $N$ of electrons,   
at different disorder strengths $W$, from $W=0$ (bottom) to 
$W/V=15$ (top) in steps of $\Delta W/V=1$. The linear system 
size is $L=10$ and the cut-off orbital $M=12$. Here and in 
the following figures, $U=-4V$ and data are averaged over 
$N_R=100$ disorder realizations.}   
\label{fig1}
\end{figure}

\begin{figure}
\begin{center}
\epsfxsize=8cm
\epsffile{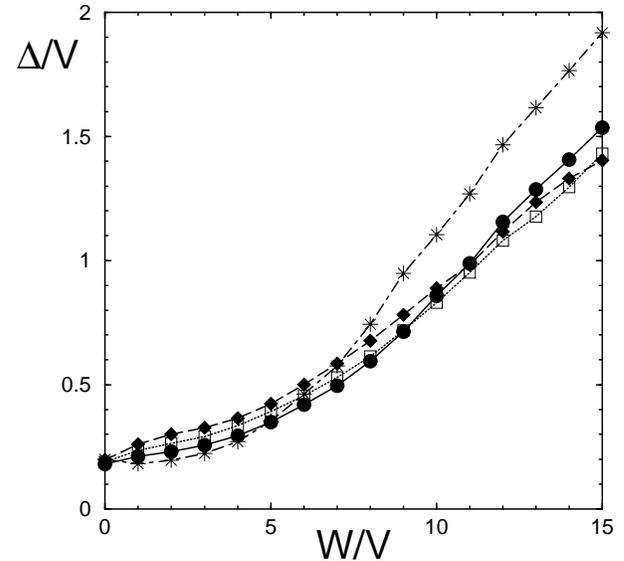}
\end{center}
\caption{Pairing gap $\Delta(N)=E_p(N)-E_p(N-1)$ vs. disorder 
strength $W$, with $E_p$ taken from Fig. \ref{fig1},  
$N=2$ (stars), $N=4$ (circles), $N=6$ (squares), and 
$N=8$ (diamonds).} 
\label{fig2}
\end{figure}

In Fig. \ref{fig1} we show the $N$-body pairing $E_p(N)$ 
as a function of the number $N$ of interacting quasiparticles 
above the Fermi sea, at different disorder strengths. 
This figure shows a clear even-odd effect, with a much 
larger increase of the pairing energy when $N$ is pair.  
This fact has a clear meaning: for $N$ even, it is possible 
to build a new pair, reducing the ground state energy due 
to the negative coupling $U<0$. For $N$ odd, in an ideal 
BCS superconductor the additional particle cannot be paired 
and remains as a quasiparticle excitation. However, a small 
ground state energy reduction is still present in our 
numerical simulations, since the unpaired particle weakly  
interacts with the superconducting pairs. 

The jump in the pairing energy from odd to even number of 
particles,  
\begin{equation} 
\label{pairing} 
\Delta(N)=E_p(N)-E_p(N-1), 
\end{equation}   
with $N$ even, can therefore be interpreted as the energy 
necessary to break a superconducting pair; in the ideal 
BCS case, this would give the superconducting energy gap. 
We note that the superconducting gap is 
extracted in a similar way in experiments with single 
Cooper pair tunneling inside superconducting islands 
\cite{tinkham,devoret}. 

In Fig. \ref{fig2} we show the pairing 
gap $\Delta(N)$ as a function of the disorder strength 
$W$, for $N=2,4,6,8$. We see that, with the exception 
of the first jump ($N=2$), the other jumps are rather 
similar. It is clear that $\Delta$ grows significantly 
with the disorder strength $W$. We attribute this effect 
to the fact that at strong disorder particles are trapped 
in the deepest minima of the random potential. 
Therefore the pair size becomes smaller that enhances 
the interaction between coupled particles, hence $\Delta$. 
  
\begin{figure}
\epsfxsize=8cm
\epsffile{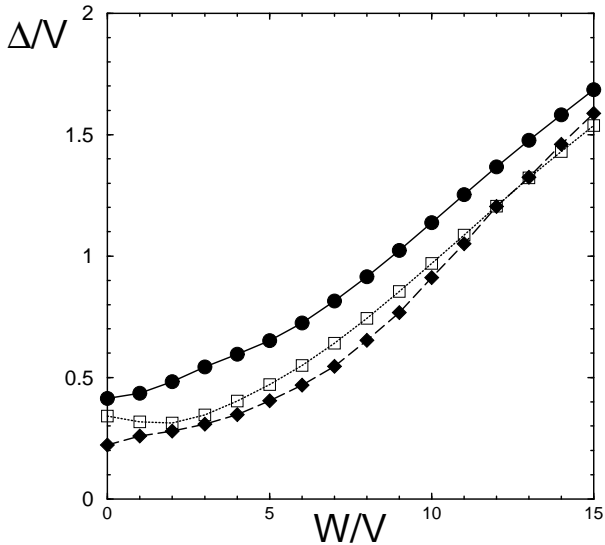}
\caption{Dependence of pairing gap $\Delta$ (average of $\Delta(N=2)$ 
and $\Delta(N=4)$) on the disorder strength
$W$, at $L=6$ (circles), $L=8$ (squares), and $L=10$ (diamonds), 
keeping the ratio $M/L^2\approx 0.2$.}  
\label{fig3}
\end{figure}

In Fig. \ref{fig3} we show the dependence of $\Delta$ on the 
system size $6\leq L \leq 10$. Since the 
Debye cut-off frequency should be independent of the 
system size, we keep constant the ratio $M/L^2\approx 0.2$.  
The coupling energy $\Delta$ becomes independent of the 
system size at large $W$, while it is not yet saturated   
at small $W$. As the pair size is determined by $1/\Delta$, 
this means that at small $W$ the size of the pair becomes 
comparable with the system size.

\section{Pair Localization} 
\label{localization} 

In order to study the localization properties of the system, 
we consider the fraction $\xi$ of the sample occupied by the 
$N$-body wavefunction $|\Psi_g\rangle$:   
\begin{equation} 
\label{fraction} 
\xi=\frac{N^2}{2L^2\sum_{{\bf i}\sigma}\rho^2_{{\bf i}\sigma}}, 
\end{equation} 
where
\begin{equation} 
\rho_{{\bf i}\sigma}= 
\langle\Psi_g|n_{{\bf i}\sigma}|\Psi_g\rangle 
\end{equation} 
is the charge density of the ground state at the site ${\bf i}$. 
With the definition (\ref{fraction}), $N/2L^2\leq\xi\leq 1$, 
the lower limit corresponding to pairs localized in a single site,
the upper limit to complete charge delocalization.   

\begin{figure}
\epsfxsize=8cm
\epsffile{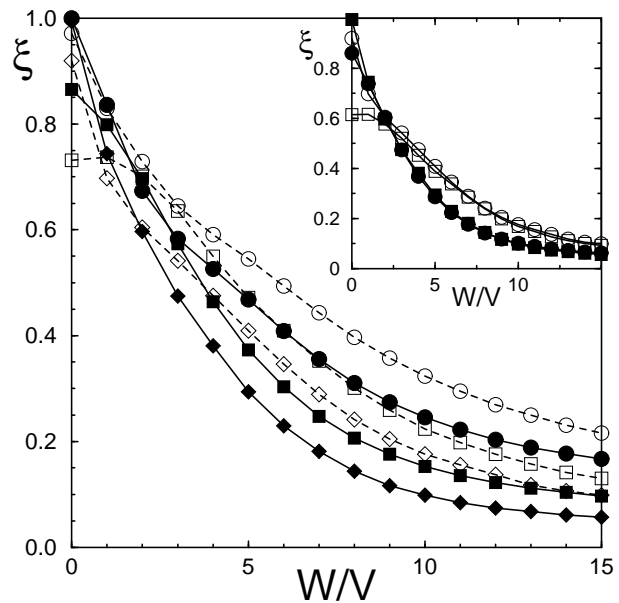}
\caption{Fraction of occupied sites $\xi$ as a function of the 
disorder strength $W$, for $M/L^2\approx 0.2$,  
$U=-4V$ (full symbols) and $U=0$ (empty symbols). 
Main figure: $N=6$ particles, 
at $L=6$ (circles), $L=8$ (squares), and $L=10$ (diamonds). 
Inset: constant particle density of mobile fermions 
$N/L^2\approx 0.06$;   
$N=4$, $L=8$ (circles) and $N=6$, $L=10$ (squares).}  
\label{fig4}
\end{figure}

In Fig.\ref{fig4} we show the fraction of occupied sites 
$\xi$ as a function of disorder, at different system 
sizes $6\leq L\leq 10$, for $N=6$ particles 
and a fixed Debye frequency ($M/L^2\approx 0.2$). 
This figure gives a clear indication of the presence of 
two regimes: at small disorder the wavefunction fills  
a large fraction of the sample (superconducting regime), 
while at large disorder $\xi$ decreases with the 
system size ($\xi\propto 1/L^2$, localized regime).      
In the inset of Fig. \ref{fig4} we show the parameter 
$\xi$ at different system sizes $L=8,10$, and  
for a constant electronic density of mobile fermions 
$N/L^2\approx 0.06$.  
The two curves puts on top of each other, suggesting  
the existence of a size-independent function $\xi(W)$ 
in the thermodynamic limit. The drop of $\xi(W)$ with $W$ 
demonstrates that disorder gives localization of Cooper pairs.

Finally we examine the question of to what extent the 
wavefunction localization is a many body effect instead 
of a single particle Anderson localization phenomenon. 
Therefore in Fig. \ref{fig4} we also show the parameter 
$\xi$ at $U=0$. The comparison between the interacting 
($U=-4$) and the noninteracting ($U=0$) case suggests    
that the interaction makes localization stronger, 
in agreement with results for two particles in a three 
dimensional random potential \cite{jose}. This effect can be 
explained qualitatively with the following argument \cite{jose}: 
attractive interaction creates pairs of total mass $m_p$ 
twice the electronic mass. This halves the effective 
hopping term $V_{\rm eff}\propto 1/m_p$, thus doubling 
the ratio $W/V_{eff}$.  
We remark that this rough argument fails in the delocalized 
regime at small disorder, where the tendency seems to be reversed. 
Actually, due to localization of single particles states 
in 2D \cite{abrahams}, for a given number $N$ of particles 
one should get $\xi(U=0)\to 0$ when $L\to\infty$. 

\begin{figure}
\epsfxsize=8cm
\epsffile{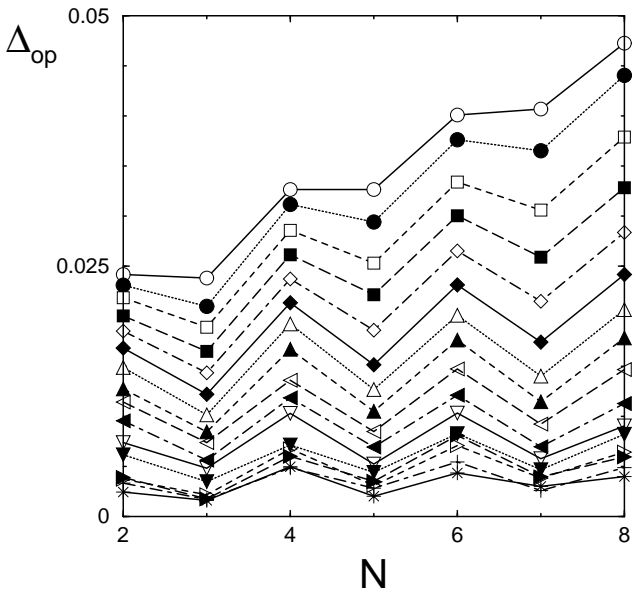}
\caption{Dependence of the order parameter $\Delta_{\rm op}$ on the number
$N$ of particles, at different disorder strengths $W$, from $W=0$ (top) 
to $W/V=15$ (bottom) in steps of $\Delta W/V=1$. 
The linear system size is $L=10$ and the cut-off orbital $M=12$.}   
\label{fig5}
\end{figure}

\section{Pair Correlation} 
\label{pair} 
 
The superconducting state can be characterized 
by the $s$-wave pair correlation function, 
\begin{equation} 
P_s({\bf r})=\langle\Psi_g|\frac{1}{L^2}\sum_{\bf i} 
\Delta_{{\bf i}+{\bf r}}\Delta_{\bf i}^\dagger|\Psi_g \rangle,     
\end{equation} 
where 
\begin{equation} 
\Delta_{\bf i}^\dagger=c_{{\bf i}\uparrow}^\dagger 
c_{{\bf i}\downarrow}^\dagger 
\end{equation} 
creates a pair at site ${\bf i}$. For an $s$-wave superconducting 
state, 
\begin{equation} 
\Delta_{\rm op}=\sqrt{P_s({\bf r}=(L/2,L/2))} 
\end{equation} 
is the order parameter of the superconductor-insulator 
transition \cite{trivedi}.   

\begin{figure}
\epsfxsize=8cm
\epsffile{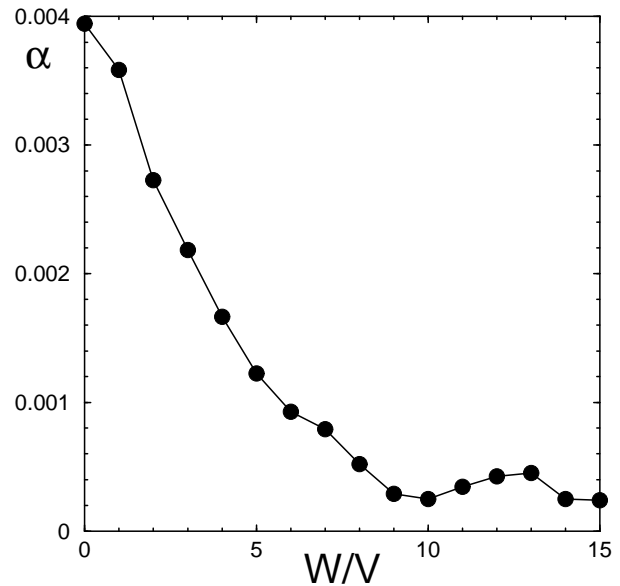}
\caption{Dependence of the slope $\alpha$ of the order parameter 
$\Delta_{\rm op}$ on the disorder strength $W$; $\alpha$ is 
extracted from a linear fit of $\Delta_{\rm op}(N)$ vs $N$, with 
data taken from Fig. \ref{fig5}.}  
\label{fig6}
\end{figure}

In Fig. \ref{fig5} we show the dependence of the order parameter 
$\Delta_{\rm op}$ on the number of interacting quasiparticles 
above the Fermi sea. 
The order parameter is strongly suppressed by disorder 
(see also Ref. \cite{trivedi}), an effect which becomes 
more evident with the addition of particles.
We remark that $\Delta_{\rm op}$ shows an approximate linear 
increase with the number of pairs. This is quite natural   
if the many body ground state wavefunction is in the BCS form 
\cite{schrieffer}, built from single particle eigenfunctions 
including disorder \cite{anderson}: 
\begin{equation} 
|\Psi_g\rangle\propto \prod_\gamma (1+g_\gamma b_\gamma^\dagger)
|0\rangle,
\end{equation}
with $b_\gamma^\dagger=d_{\gamma\uparrow}^\dagger 
d_{\gamma\downarrow}^\dagger$, and $g_\gamma$ variational parameters. 
Using the relation $c_{{\bf i}\sigma}^\dagger=\sum_\alpha 
\phi_\alpha({\bf i}) d_{\alpha\sigma}^\dagger$, after lattice and 
disorder averaging the dominant contributions in $P_s({\bf r})$ 
is proportional to the number of pairs, 
$\Delta_{\rm op}(N)\propto \alpha\frac{N}{2}$ 
(here we have taken into account that $g_\gamma$ in the BCS 
theory changes smoothly with $\alpha$ around the Fermi level 
\cite{schrieffer} and here we considered $N\ll 2n_F$).
Here $\alpha$ is a parameter which determines the slope of
$\Delta_{\rm op}$ variation with $N$.
The order parameter decreases when an unpaired particle is 
added. In our opinion, this is due to the fact that this extra 
electron weakly interacts with the paired particles, reducing 
the pair correlation function. 

In Fig.\ref{fig6} we show the slope $\alpha$ of the linear 
fit of the order parameter $\Delta_{\rm op}$ as a function 
of the number $N$ of quasiparticles.  
The suppression of this quantity with disorder is evident, 
indicating a rather sharp crossover from a superconducting 
to an insulating behavior in our finite size lattice.  

\section{Conclusions} 
\label{conc} 

In this paper we have investigated the localization of 
Cooper pairs for small clusters in a two-dimensional disordered 
substrate. We have shown that the Cooper pair coupling energy  
displays an even-odd asymmetry: this parity effect survives also 
in the presence of disorder. The pairing gap is strongly enhanced 
by disorder, which at the same time leads to localization of
Cooper pairs (gapped insulator). Therefore, in the insulating 
regime, the breaking of Cooper pairs should enhance transport.  
This is consistent with the resistivity drop observed in 
experiments with an applied magnetic field 
\cite{paalanen,gantmakher}, which might signal the 
crossover from a Cooper pair insulator to an 
electronic insulator.


\begin{thebibliography}{99}
\bibitem{exp} For a review see, e.g., A.M. Goldman and 
N. Markovi\'c, Physics Today {\bf 51}, November Issue, 39 (1998). 
\bibitem{fisher} M.P.A. Fisher, G. Grinstein, and S.M. Girvin, 
Phys. Rev. Lett. {\bf 64}, 587 (1990).
\bibitem{erik} M. Wallin, E. S. S\o rensen, S. M. Girvin, 
               and A. P. Young, Phys. Rev. B {\bf 49}, 12115 (1994).
\bibitem{trivedi} R.T. Scalettar, N. Trivedi, and C. Huscroft, 
Phys. Rev. B {\bf 59}, 4364 (1999). 
\bibitem{trivedi2} A. Ghosal, M. Randeria, and N. Trivedi, 
Phys. Rev. Lett. {\bf 81}, 3940 (1998).  
\bibitem{ralph} C.T. Black, D.C. Ralph, and M. Tinkham, 
Phys. Rev. Lett. {\bf 76}, 688 (1996). 
\bibitem{delft} J. von Delft, A.D. Zaikin, D.S. Golubev, 
and W. Tichy, Phys. Rev. Lett. {\bf 77}, 3189 (1996). 
\bibitem{ambegaokar} R.A. Smith and V. Ambegaokar, 
Phys. Rev. Lett. {\bf 77}, 4962 (1996). 
\bibitem{larkin} K.A. Matveev and A.I. Larkin, 
Phys Rev. Lett. {\bf 78}, 3749 (1997).  
\bibitem{cooper} L.N. Cooper, Phys. Rev. {\bf 104}, 1189 (1956). 
\bibitem{jose} J. Lages and D.L. Shepelyansky, Phys. Rev. B
{\bf 62}, 8665 (2000). 
\bibitem{abrahams} E. Abrahams, P.W. Anderson, D.C. Licciardello, and 
T.V. Ramakrishnan, Phys. Rev. Lett. {\bf 42}, 673 (1979).  
\bibitem{dagotto} R.T. Scalettar, E.Y. Loh, J.E. Gubernatis,
A. Moreo, S.R. White, D.J. Scalapino, R.L. Sugar, and 
E. Dagotto, Phys. Rev. Lett. {\bf 62}, 1407 (1989).   
\bibitem{sorensen} D.C. Sorensen, SIAM J. Mat. Anal. Appl. 
{\bf 13}, 357 (1992). 
\bibitem{tinkham} M.T. Tuominen, J.M. Hergenrother, T.S. Tighe, 
and M. Tinkham, Phys. Rev. Lett. {\bf 69}, 1997 (1992). 
\bibitem{devoret} M.H. Devoret, D. Esteve, and C. Urbina, 
in {\it Mesoscopic Quantum Physics}, Les Houches Session 
LXI, edited by E. Akkermans, G. Montambaux, J.-L. Pichard, 
and J. Zinn-Justin (Elsevier, Amsterdam, 1995).  
\bibitem{schrieffer} J.R. Schrieffer, {\it Theory of 
Superconductivity}, Perseus Books (Reading, Massachusetts, 1983). 
\bibitem{anderson} P.W. Anderson, J. Phys. Chem. Solids 
{\bf 11}, 26 (1959). 
\bibitem{paalanen} M.A. Paalanen, A.F. Hebard, and R.R. Ruel, 
Phys. Rev. Lett. {\bf 69}, 1604 (1992). 
\bibitem{gantmakher} V.F. Gantmakher, M.V. Golubkov, 
V.T. Dolgopolov, A.A. Shashkin, and G.E. Tsydynzhapov, 
cond-mat/0004377.  
\end{thebibliography}
\end{document}